# Enhancing Selective Encryption for H.264/AVC Using Advanced Encryption Standard

M. Abomhara, Omar Zakaria, Othman O. Khalifa, A.A Zaidan, B.B Zaidan.

*Abstract*—Multimedia information availability has increased dramatically with the advent of mobile devices. But with this availability comes problems of maintaining the security of information that is displayed in public. Many approaches have been used or proposed for providing security for information dissemination over networks, protection system classified with more specific as encryption information, and combination between video compression and encryption to increase information security. The strength of the combination between Video compression and encryption science is due to the non-existence of standard algorithms to be used in video compression and encrypting secret video stream. Also there are many ways could be used in video encryption methods such as combining several different encryption methods with video compression to pass a secret video streaming. Furthermore, there is no formal method to be followed to discover an encrypted video. For this reason, the task of this paper becomes difficult. In this paper proposed a new system of video encryption is presented. The proposed system aim to gain a deep understanding of video data security on multimedia technologies, to investigate how encryption and decryption could be implemented for real time video applications, and to enhance the selective encryption for H.264/AVC. The system includes two main functions; first is the encoding/encryption of video stream, through the execution of two processes (the input sequences of video is first compressed by the H.264/AVC encoder, and the encoded bit stream (I-frame) is partially encrypted using AES block cipher). And the second function is the decryption/decoding of the encrypted video through two process (specify the encrypted I-frame stream, decryption of the I-frame, and decoding with H.264/AVC decoder). The system is implemented by using Matlab.

*Index Terms*—Encryption /Decryption, H.264/AVC, Video Compression. Advance Encryption Standard (AES)

I. INTRODUCTION

With the rapid growth of Internet and multimedia applications in distributed environments, it becomes easier for digital data owners to transfer multimedia documents across all over the world via the Internet. Therefore, multimedia security [1] has become one of the most aspects of communications with the continuous increase in the use of digital data transmission. In addition, some applications, such as TV broadcast video in demand and video conferencing require a special and reliable secure storage or transmission of digital images and videos which may use in many applications. Multimedia security in general is provided by a method or a set of methods used to protect the multimedia contents. These methods where heavily based on cryptography.[1][2] However, cryptography is the art of keeping information secret by transforming it into an unreadable format (encryption) by using special keys, then rendering the information readable again for trusted parties by using the same or other special keys (decryption)[3]. Moreover, modern cryptography does not confine itself to only maintaining the secrecy of information but goes beyond that by ensuring the identity of communicating parties (authentication), ensuring that information has not been tampered with others (integrity), and preventing that any of the communicating parties denies having received or sent information (non-repudiation). In Multimedia data, cryptography is necessary when communicating over any untrusted medium including public networks particularly the Internet. In addition to protect information from theft, alteration or misuse, cryptography can also be used for user authentication. In modern field of cryptography there are three types of cryptographic schemes: secret key (or symmetric) cryptography, public-key (or asymmetric) cryptography, and hash functions [3]. Symmetric-key cryptography refers to encryption methods in which both the sender and receiver share the same key. This was the only kind of encryption publicly known until 1976 when White Diffie and Martin Hellman proposed the notion of public-key cryptography [4][5], which, also known as public key cryptography, it uses two keys, one for encrypting information by sender, and one for decrypting information by the receiver. A Hash function, also called message digest or one-way encryption, it is a transformation that takes an input and returns a fixed-size string, which is called the hash value. While symmetric or asymmetric encryption methods provide means to ensure information confidentiality, hash functions, provide a measure of the integrity of a file. For instance hash algorithms are typically used to provide a digital signature of a file's contents often used to ensure that the file has not been altered by an intruder or virus. In some applications such as commercial TV broadcast, military applications [6][7], and intelligence has to have a secured transmission or storage media. Furthermore,

Manuscript received June 20, 2009
Mohamed Abomhara, - Masters Student, Department of Computer Science & Information Technology, University Malaya, Kuala Lumpur, Malaysia, and phone: +60172471620 Email: m.abomhara@gmail.com.
Dr.Omar Zakaria - Senior lecturer, Department of Computer Science & Information Technology, University Malaya, Kuala Lumpur, Malaysia,omar@fsktm.um.edu.my.
Dr. Othman O. Khalifa - Head of the department of Electrical and Computer Engineering, International Islamic University Malaysia., Kuala Lumpur, Malaysia,
Aos Alaa Zaidan – PhD candidate, Department of Electrical & Computer Engineering, International Islamic University Malaya, Kuala Lumpur, Malaysia, phone:+6012452457 Email:aws.alaa@yahoo.com.
Bilal Bahaa Zaidan - PhD candidate, Department of Electrical & Computer Engineering, International Islamic University Malaya, Kuala Lumpur, Malaysia, Email: bilal_bahaa@hotmail.com

videoconferencing has become a daily characteristic of financial businesses. As it saves time, effort, and travel expenses for large companies [8]. This communicated video application has to be completely secured against theft, alteration or misuse. For this purpose, cryptography algorithms can be applied to these multimedia applications to ensure their security. Communication security can be accomplished by means of standard symmetric key cryptography such media can be treated as binary sequence and the whole data can be encrypted using a cryptosystem such as AES or DES [9]. In general, when the multimedia data is static (not a real-time streaming) we can treat it as a regular binary data and use the conventional encryption techniques. Previously, encryption the entire video data using standard encryption algorithms, it is referred as Naive approach, this method can provide substantial high security but it needs huge computational cost. At this time, most of researches are about selective video data encryption, which can reduce computational cost as it just encrypts only a part of video data. Due to variety of constraints (such as the near real-time speed, etc.), communication security for streaming audio and video media is harder to accomplish. Encryption of video and audio multimedia content is not simply the application of established encryption algorithms, such as DES or AES, to its binary sequence. It involves careful analysis to determine and identify the optimal encryption method when dealing with video data. Recently, encryption techniques provide the basic technology for building secure multimedia system. In order to provide real time reliable security of digital images and videos, many different encryption algorithms have been brought forward to secure networked continuous media from potential threats such as hacker and eavesdroppers and most of video encryption algorithms are designed for various video coding standards such as MPEG-1, MPEG-2/H.262, and MPEG-4 [8][10]. Unfortunately, these encryption algorithms do not appropriate for secure the current multimedia. Therefore, current research is focused on modifying and optimizing the existing cryptosystems for real-time video. It is also oriented towards exploiting the format specific properties of many standard video formats in order to achieve the desired speed and enable real-time security streaming.

## II. MOTIVATION AND SIGNIFICANT

The use of multimedia has increased significantly among consumers around the world. The fact that multimedia is becoming a part of the human life. With the absence of the reliable security to protect multimedia data, there is a risk for multimedia users specially when dealing with sensitive information over public network like Internet. it is necessary to provide adequate security for it, so that the service can be a reliable tool for financial transactions. The need for end-to-end encryption for multimedia data, since communication between users is encrypted but the data is not encrypted while traveling through a public network or saved in one place, necessitates finding a solution to provide end-to-end encryption to multimedia contents. The exponential growth of security incidents and constant attempts by hackers to compromise confidentiality and integrity underline the necessity for stronger security for multimedia contents specially when used to send/receive confidential information or used to perform financial transactions. Some solutions are already offered to encrypt multimedia data using symmetric key algorithm with MPEG bit stream, so symmetric key algorithm solutions to multimedia data with H.264/AVC is natural evolution to that solution to provide end-to-end encryption for multimedia.

## III. H.264/AVC AND SELECTIVE ENCRYPTION

### A. H.264/AVC

H.264 is an industry standard for video compression, the process of converting digital video into a format has low bitrate capacity when it is stored or transmitted. The main objective behind the H.264[11][12] development was to build up a high performance video coding standard by adopting a back to basics approach with simple and straightforward design using well known blocks. H.264/AVC based on conventional block based motion-compensated video coding as same as the exiting standards, but with a numbers of new features and advantages significantly improve its rate-distortion performance and distinguish it from the exiting standards such as MPEG-2, MPEG-4 Part 2, H.263 however, at the same time, sharing common features with the exiting standards. The main features that improve coding efficiency are the following [10].

- Variable block-size motion compensation with the block size as small as 4x4 pixels.
- Quarter-sample motion vector accuracy.
- Motion vectors over picture boundaries.
- Multiple reference picture motion compensation.
- In-the-loop deblocking filtering.
- Small block-size transformation (4x4 block transform).
- Enhanced entropy coding methods (Context-Adaptive Variable-Length Coding (CAVLC) and Context-Adaptive Binary Arithmetic Coding (CABAC)).

Special attention is given to the improvement of working with the losses during transmission over different networks or robustness to data errors [13]. The H.264/AVC consists of two conceptual layers [12][14]. The Video Coding Layer (VCL) which is efficiently represents the content of the video data, and the Network Abstraction layer (NAL) that is designed to convert the VLC representation into format suitable for specific transport layers or storage media. Some key concepts of the NAL are NAL unites. Each NAL unit is effectively a packet that contains an integer number of bytes, including a header and payload. The header byte is the first byte of each NAL unit that contains an indication of the type of data, and the remaining byte is the payload that represents the related data. A NAL unit specifies a generic format for use in both packet-oriented and bite stream system, as well as the NAL units are classified into VCL unites which contain the data that represents the values of the samples in the video pictures, and the non-VCL NAL units which contain any associated additional information such as parameter sets. The structure of H.264/AVC is shown in fig. 1.

Since H.264/AVC is designed to address a large range of applications, the structure of its bitstream may vary significantly. To avoid implementing of all possible stream structures by each specification-conform decoder, profiles was defined. A profile is a subset of the capabilities including the entire bitstream syntax. H.264/AVC standard includes the following three profiles, targeting specific classes of applications:

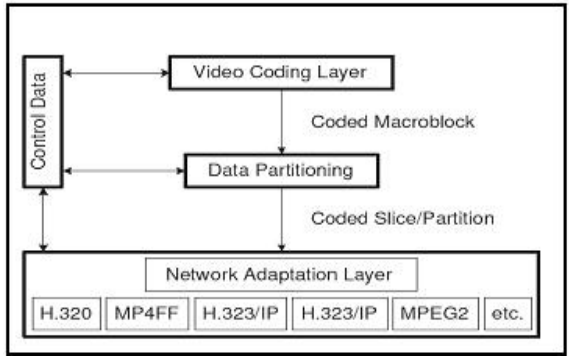

Figure .1 Structure of H.264/AVC Video Encoder

- Baseline Profile: Primarily for lower-cost applications demanding less computing resources, this profile is used widely in video conferencing and mobile applications.(Note that baseline profile does not support B slices, only I and P slices are possible).
- Main Profile: Originally intended as the mainstream consumer profile for broadcast and storage applications, the importance of this profile faded when the High profile was developed for those applications.
- Extended Profile: Intended as the streaming video profile, this profile has relatively high compression capability and some extra tricks for robustness to data losses and server stream switching.

Moreover, H.264/AVC uses the two entropy coding method, termed Context-adaptive variable-length coding(CAVLC) and Context-adaptive binary arithmetic coding(CABAC). CAVLC is an algorithm with a lower-complexity for the coding of quantized transform coefficient values. And, CABAC [15][16] is an algorithm to lossless compress syntax elements in the video stream knowing the probabilities of syntax elements in a given context. CABAC compresses data more efficiently than CAVLC but requires considerably more processing to decode. Thus, CAVLC is used in the base-layer mainly and CABAC is used optionally in the enhancement-layer.

IV. SELECTIVE ENCRYPTION

Selective encryption is a technique of encrypting some parts of a compressed data file while send-off others unencrypted. It is a strategy that small fraction of encrypted bits can reason a high ratio of damage to a file. Instead of encrypting the whole file bit by bit, only highly sensitive bits are changed as shown in fig. 2 [6]. Moreover selective encryption reduces required total encryption work and saves system resources as it just encrypts some part of video stream for example The basic selective encryption is based on the H.264/AVC I-frame, P-frame, and B-frame structure. It encrypts the I-frame only because, conceptually P- and B-frame are useless without knowing the corresponding I-frame[7][8][17]. We propose a technique that selectively encrypts some parts of compressed video file while guarantee the security of the original file. We reduce the time for encrypting video file, but also system complexity. The idea of this scheme is to encrypt different levels of selective parts of H.264/AVC stream by using the feature of H.264/AVC layered structures.

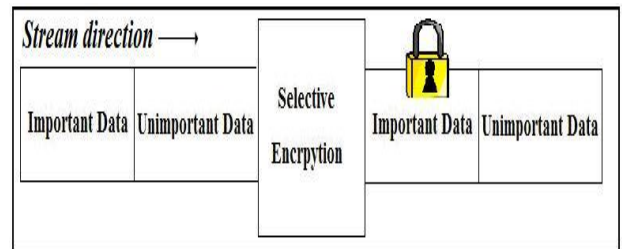

Figure .2 Structure of Selective Encryption.

V. ADVANCED ENCRYPTION STANDARD (AES)

In 1997, the NIST called for submissions for a new standard to replace the aging DES. The contest terminated in November 2001with the selection of the Rijndael cryptosystem as the Advanced Encryption Standard (AES) [3][9]. The Rijndael cryptosystem operates on 128-bit blocks, arranged as $4 \times 4$ matrices with 8-bit entries. The algorithm can use a variable block length and key length. The latest specification allowed any combination of keys lengths of 128, 192, or 256 bits and blocks of length 128, 192, or 256 bits.

AES may, as all algorithms, be used in different ways to perform encryption. Different methods are suitable for different situations. It is vital that the correct method is applied in the correct manner to each and every situation, or the result may well be insecure even if AES as such is secure. It is very easy to implement a system using AES as its encryption algorithm, but much more skill and experience are required to do it in the right way for a given situation. To describe exactly how to apply AES for varying purposes is very much out of scope for this paper.

A. Strong keys

Encryption with AES is based on a secret key with 128, 192 or 256 bits. But if the key is easy to guess it doesn't matter if AES is secure, so it is as critically vital to use good and strong keys as it is to apply AES properly. Creating good and strong keys is a surprisingly difficult problem and requires careful design when done with a computer. The challenge is that computers are notoriously deterministic, but what is required of a good and strong key is the opposite unpredictability and randomness. Keys derived into a fixed length suitable for the encryption algorithm from passwords or pass phrases typed by a human will seldom correspond to 128 bits much less 256. To even approach 128- bit equivalence in a pass phrase, at least 10 typical passwords of the kind frequently used in day-to-day work are needed.

Weak keys can be somewhat strengthened by special techniques by adding computationally intensive steps which increase the amount of computation necessary to break it. The risks of incorrect usage, implementation and weak keys are in no way unique for AES, these are shared by all encryption algorithms. Provided that the implementation is correct, the security provided reduces to a relatively simple question about how many bits the chosen key, password or pass phrase really corresponds to. Unfortunately this estimate is somewhat difficult to calculate, when the key is not generated by a true random generator.

*B. The Round Transformations*

There are four transformations:
- Add Round Key

Add Round Key is an XOR between the state and the round key. This transformation is its own inverse.

- Sub Bytes

Sub Bytes is a substitution of each byte in the block independent of the position in the state. This is an S-box. It is bisection on all possible byte values and therefore invertible (the inverse S-box can easily be constructed from the S-box). This is the non-linear transformation. The S-box used is proved to be optimal with regards to non-linearity. The S-box is based on arithmetic in GF ($2^8$).

- Shift Rows

Shift Rows is a cyclic shift of the bytes in the rows in the state and is clearly invertible (by a shift in the opposite direction by the same amount).

- Mix Columns

Each column in the state is considered a polynomial with the byte values as coefficients. The columns are transformed independently by multiplication with a special polynomial $c(x)$. $c(x)$ has an inverse $d(x)$ that is used to reverse the multiplication by $c(x)$.

*C. The Rounds*

A round transformation is composed of four different transformations as shown in fig.3.

The Round keys are made by expanding the encryption key into an array holding the Round Keys one after another. The expansion works on words of four bytes. Nk is a constant defined as the number of four bytes words in the key. The encryption key is filled into the first Nk words and the rest of the key material is defined recursively from preceding words. The word in position i, W[i], except the first word of a Round Key, is defined as the XOR between the preceding word, W[i-1], and W[i-Nk]. The first word of each Round Key, W[i] (where i mod Nk == 0), is defined as the XOR of a transformation on the preceding word, T (W [i - 1]) and W [i - Nk]. The transformation T on a word, w, is w rotated to the left by one byte, XOR'ed by a round constant and with each byte substituted by the S-box.

```
Round (State, RoundKey) {
    SubBytes(State);
    ShiftRows(State);
    MixColumns(State);
    AddRoundKey(State, RoundKey);
}
```
Figure 3: Four Different Transformations.

The final round is like a regular round, but without the mix columns transformation as shown in fig. 4:

```
FinalRound(State, RoundKey) {
    SubBytes(State);
    ShiftRows(State);
    AddRoundKey(State, RoundKey);
}
```
Figure 4: Final Round.

VI. METHODOLOGY

*A. Proposed System Overview*

Selective encryption is a technique for encrypting only parts of a compressed video stream to reduce computational complexity Selective encryption is not a new idea. It has been proposed in several applications, especially in multimedia system. Selective encryption can be used to reduce the power consumed by the encryption function for digital content. Since particular parts of the bit stream are encrypted, selective encryption can also enable new system functionality such allowing previewing of content. For selective encryption to work, we need to rely on a characteristic of the compression algorithm to concentrate important data relative to the original signal in a relatively small fraction of the compressed bitstream. These important components of the compressed data become candidates for selective encryption. In our selective encryption, an I-Frame bit stream of H.264/AVC, bitstream is encrypted to minimize computational complexity or provide new functionalities for uses of the encrypted bit stream while at the same time providing reasonable security of the bit stream. The block diagram of our proposed selective encryption method for video compressed, using the H.264/AVC video coding method is shown in Figure 6. The input video is first compressed by the H.264/AVC encoder. The output bitstream of the H.264/AVC encoder consists of individual types of data, the video frames (pixels), Intra frame, inter frames, and etc. The encoded bit stream (I-frame) is partially encrypted using AES block cipher with key size of 128 bits, which is XOR-ed with the cipher key to generate the cipher data. Keeping the rest of the data unencrypted because we believe that encrypting I-frame only is more signification. Due to the fact that conceptually P- and B- frame are useless without knowing the corresponding I-frame

## B. Concept of proposed System

The idea of the proposed system is to apply a cipher that will alter the video stream only visually while keeping it fully decodable. As showed in Figure 5. A way to determine which bits can be ciphered could be to test over the a set of sequences in which parts in the bitstream bit inversion can be applied without crashing the decoder process, and brining only visual errors. Obviously this is only a first approach that must be followed by a careful study of the standard and the requirements to determine all or many of the possibilities to encrypt bits without leading to a potentially non-compatible stream.

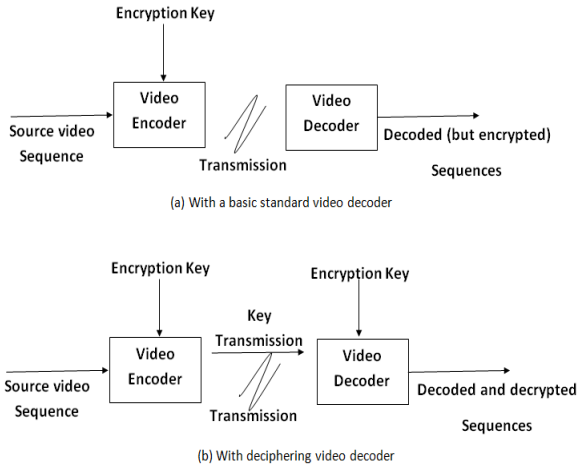

Figure 5. Principle of coding/transmission/decoding chain with encryption mechanism for decrypting and not-decrypting video decoder

In this proposed system, the bits to be encrypted are chosen with respect to the considered video standard to ensure full compatibility, achieved by selecting the bit (I-Frame bitStream) for which each the encrypted configuration modify negligibly the decoding process contexts in sense where their introduction does not create de-synchronization nor lead to non-compliant bitstream. As such, an encryption operation leading to a change of symbol table used in the coding process is not negligible whereas an encryption operation that leads to interpreting a given I-Frame bits. In each case, it is important to note that the bits should maintain this capacity in every coded bitstream, and that it can not be envisaged to consider cases where given configuration of bitstream will allow immediate or delayed resynchronization. The interest to choose carefully the way to encryption is performed is double: 1) ensures the compatibility with the requirements of the considered video standard. 2) Makes it difficult for cryptanalysis attacks to find an angle to break the encryption key, as it is aimed at making all solutions possible, hence removing possibilities to rule out some cases based on non-respect of standard syntax.

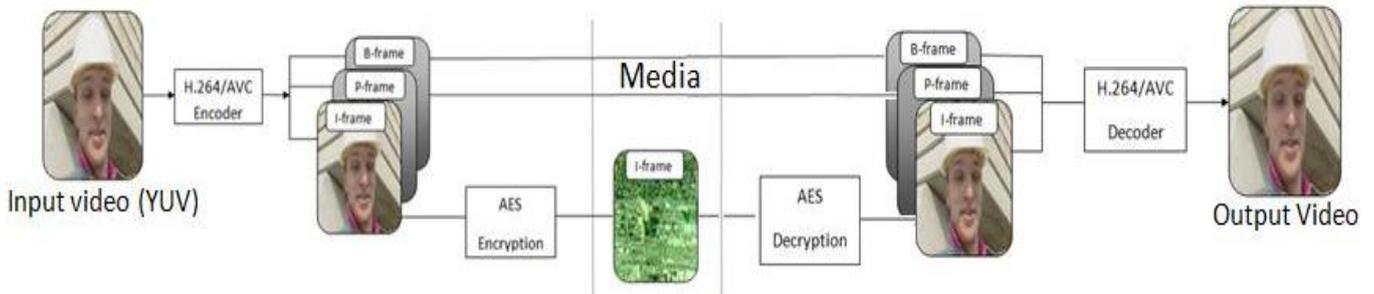

Figure 6. Diagram of the Proposed Selective Encryption Method

## C. The Proposed Scheme Structure

To protect the video streaming information from against theft, alteration or misuse before transmission or storage our system encrypts the output of the H.264/AVC bitstream (I-frame bitstream) by the applied encryption algorithm (AES) which is mentioned early. The block diagram for the operation can be performed as shown in Figure 7. The block flow retraction operation can be performed as shown in Figure 8.

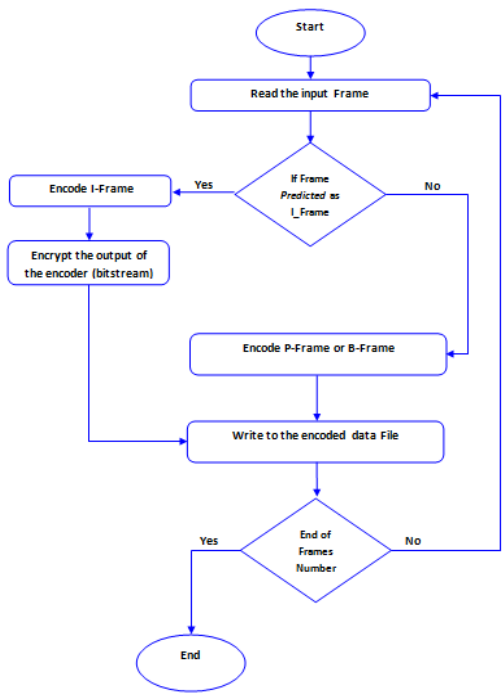

Figure 7. The Block Diagram of Encoder/Encryption Function

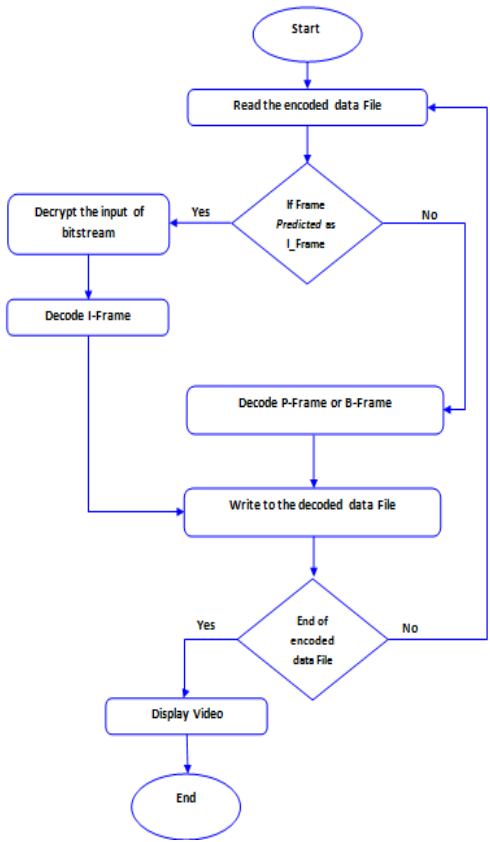

Figure 8. The Block Diagram of Decryption/Decoder Function

## VII. CONCLUSION AND DISCUSSION

This paper a proposed system to study video data security by a review and investigation on multimedia security technologies and successful to achievement main above goals, which enhancing the Selective Video Encryption Using Computation between H.264/AVC and AES Encryption Algorithm

H.264/AVC bit-stream was able to be consumed in a form of the partial bit-stream converted by extraction process. At this time, each bitsrtream of the H.264/AVC bitstream should be consumed by a trusted user who has the rights for accessibility. The requirements for H.264/AVC encryption and decryption are as follow.

- For the video encryption

The encryption considers the security, time efficiency, format compliance and error robustness, basically. The I-Frame encryption being the basis of P-frame and B-Frame in H.264/AVC bit-streams should be more securely protected than the enhancement layers.

- For the video decryption

The quality of decrypted contents has to be degraded than the quality of original contents if the given bit-stream is decrypted with only the layer keys of a lower level than the given bitstream level.

However, these requirements have trade-off between the security and time efficiency in the encryption and decryption process. The importance of encryption system is to encrypts the intra prediction modes. Therefore, the proposed encryption system provides the low computational complexity, low bit-overhead, and format compliance in encoding basically. Moreover, it offers the coding efficiency through the selective encryption and the security through the use of same keys as well as the sufficient quality degradation of video content. The proposed System implemented by matlab.

ACKNOWLEDGEMENT

Thanks in advance for the entire worker in this project, and the people who support in any way, also I want to thank UM and IIUM for the support they offered.

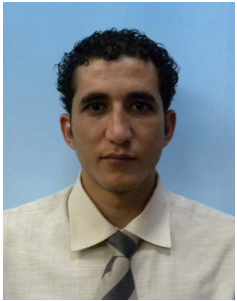

**Mohamed Abomhara-**. He completed his undergraduate degree in Computer Science at 7th October University Banewiled, Libya in 2006. He is a Master candidate in Data Communication and Computer Network at Faculty of Computer Science & Information Technology University of Malaya, Malaysia. He has done many projects on video encryption. His research interests are in computer security, encryption/decryption and video processing.

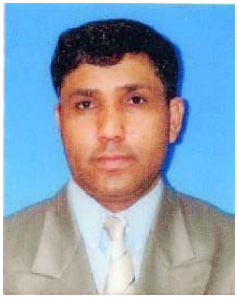

**Othman O. Khalifa** received his Bachelor's degree in Electronic Engineering from the Garyounis University, Libya in 1986. He obtained his Master degree in Electronics Science Engineering and PhD in Digital Image Processing from Newcastle University, UK in 1996 and 2000 respectively. He worked in industrial for eight years and he is currently Professor and Head of the department of Electrical and Computer Engineering, International Islamic University Malaysia. His area of research interest is Communication Systems, Information theory and Coding, Digital image / video processing, coding and Compression, Wavelets, Fractal and Pattern Recognition. He published more than 130 papers in international journals and Conferences. He is SIEEE member, IEEE computer, Image processing and Communication Society member.

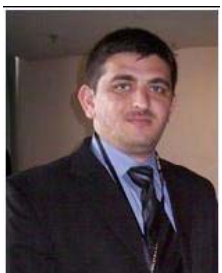

**Aos Alaa Zaidan** - He obtained his 1st Class Bachelor degree in Computer Engineering from university of Technology / Baghdad followed by master in data communication and computer network from University of Malaya. He led or member for many funded research projects and He has published more than 40 papers at various international and national conferences and journals, he has done many projects on Steganography for data hidden through different multimedia carriers image, video, audio, text, and non multimedia carrier unused area within .exe file, Cryptography and Stego-Analysis systems, currently he is working on the multi module for Steganography, Development & Implement a novel Skin Detector, he is also interest in the three-dimensional biometrics. He is PhD candidate on the Department of Electrical & Computer Engineering / Faculty of Engineering / International Islamic University Malaya / Kuala Lumpur/Malaysia. He is members IAENG, CSTA, WASET, and IACSIT. He is reviewer in the IJSIS, IJCSNS, IJCSN and IJCSE.

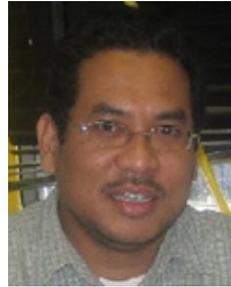

**Omar Zakaria-** He completed his undergraduate degree in Computer Science at the Computer Centre, University of Malaya (UM), Kuala Lumpur in 1994. He started work as an analyst programmer in Maybank Bhd at Maybank Tower, Jln Tun Perak, Kuala Lumpur in October 1994. However, he left Maybank in February 1995 because he got scholarship from UM to pursue his Master degree. He obtained his Master and PhD in information systems security management from the Royal Holloway, University of London, United Kingdom, in 1996 and 2007, respectively. He joined the University of Malaya as a tutor in February 1995 at Centre for Foundation Studies in Science. He was appointed to Lecturer in December 1996, and transferred to Faculty of Computer Science & Information Technology in April 1997. Subsequently, he was promoted to Senior Lecturer in April 2006. His research interests are in information systems security management (ISMS).

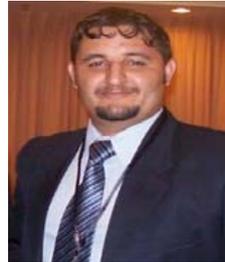

**Bilal Bahaa Zaidan** - has received bachelor degree from Saddam University in the Mathematics and Computer Application, Baghdad-Iraq fallowed by master from Department of Computer System & Technology/ Faculty of Computer Science and Information Technology/University of Malaya /Kuala Lumpur/Malaysia, He led or member for many funded research projects and He has published more than 40 papers at various international and national conferences and journals. His research interest on Steganography & Cryptography with his group he has published many papers on data hidden through different multimedia carriers such as image, video, audio, text, and non multimedia careers such as unused area within .exe file, he has done projects on Stego-Analysis systems, currently he is working on Quantum Key Distribution QKD and multi module for Steganography, he is also interest in the three-dimensional biometrics.  He is PhD candidate on the Department of Electrical & Computer Engineering / Faculty of Engineering / International Islamic University Malaya /Kuala Lumpur/Malaysia. He is members IAENG, CSTA, WASET, and IACSIT. He is reviewer in the IJCEE, IJCSNS, IJCSN and IJCSE.